\documentclass[prd,notitlepage]{revtex4-1}

\usepackage{graphicx}
\usepackage{amsmath,amssymb}
\usepackage{bm}
\usepackage{hyperref}

\hypersetup{
    colorlinks=true,
    linkcolor=red,
    citecolor=blue,
}

\def\be{\begin{equation}}
\def\ee{\end{equation}}
\def\ba{\begin{eqnarray}}
\def\ea{\end{eqnarray}}

\begin{document}

\title{
Cosmological magnetic field correlators from blazar induced cascade
}

\date{\today}

\author{Hiroyuki Tashiro and Tanmay Vachaspati}
\affiliation{ 
Physics Department, Arizona State University, Tempe, Arizona 85287, USA.
}

\begin{abstract}
TeV blazars offer an exciting prospect for discovering cosmological magnetic
fields and for probing high energy processes, including CP violation, in the early universe. We propose a method for reconstructing both the non-helical and the 
helical magnetic field correlators using observations of cascade 
photons from TeV blazars. 
\end{abstract}


\maketitle

\section{Introduction}
\label{introduction}

Large-scale magnetic fields with micro Gauss strength have been observed
in galaxies~\cite{1992ApJ...387..528K,Bernet:2008qp} and clusters of
galaxies~\cite{Clarke:2000bz,Bonafede:2010xg}.
Although it is assumed that such
magnetic fields are the result of amplification from weak seed
magnetic fields, the origin of the seed magnetic field has not yet been understood.
There are two classes of models for the seed generation mechanism:
astrophysical and cosmological models.  In astrophysical
models, the seeds are associated with nonlinear structures, and are
generated during structure formation 
through the Biermann battery effect~\cite{Biermann:2003xh}, or the Weibel
instability~\cite{2009ApJ...693.1133L}. 
On the other hand, seeds in the cosmological models
are produced in the early universe and exist as extragalactic magnetic fields.
This class of models includes the generation mechanisms in an inflationary epoch, during
cosmological phase transitions, and at the epoch of recombination (for recent reviews, see~\cite{Kandus:2010nw,Widrow:2011hs,Durrer:2013pga}).

Recent gamma ray observations suggest the existence of cosmological
magnetic fields stronger than $\sim 10^{-16}$ Gauss in the
voids~\cite{Neronov:1900zz, Tavecchio:2010mk,Dolag:2010ni}.  Although
more detailed work will be required to establish this lower
limit~\cite{Broderick:2011av,Miniati:2012ge,2012ApJ...758..102S},
the suggestion of such
magnetic fields in large-scale structure voids, $\sim 100~{\rm Mpc}$ away from 
non-linear structures, provides a strong argument in favor of cosmological models.
Further support in favor of an early universe origin can be obtained
if the cosmological galactic magnetic fields are ``helical'', thus indicating a process of
magneto-genesis that fundamentally violates invariance under charge conjugation
plus parity reflection (CP)~\cite{Vachaspati:2001nb}. 

Magnetic helicity density is defined as
\begin{equation}
 h = {1 \over V} \int_V  d^3 x ~ {\bm A} \cdot  {\bm B},
\end{equation}
where $\bm A$ is the vector potential of magnetic fields ${\bm B}$ with
${\bm B } = \nabla \times {\bm A}$. Magnetic helicity is odd under CP transformations
as ${\bm A}$ and ${\bm B}$ are odd under C, while ${\bm A}$ is even but
${\bm B}$ is odd under P. 
Non-zero magnetic helicity is predicted in scenarios in which cosmic baryogenesis 
and magneto-genesis occur concurrently during a cosmological phase
transition~\cite{Vachaspati:1991nm,Joyce:1997uy,Cornwall:1997ms,Vachaspati:2001nb,Tashiro:2012mf}.
Then the magnetic helicity density is related to the cosmic baryon number density
and the CP violation responsible for the excess of matter over antimatter also provides
helicity to the magnetic field. Other scenarios that can generate helical magnetic fields 
that are not tied to baryogenesis have been studied in Refs.~\cite{Field:1998hi,Campanelli:2005ye,Campanelli:2008kh}.  Helicity is also
an important factor in the evolution of magnetic fields because it helps to transfer
magnetic field energy from small to large length scales, a process called an 
``inverse cascade''.  Due to the inverse cascade, helical magnetic fields can grow to
astrophysically relevant scales at the present epoch, even though the initial scale 
of the magnetic field extends only up to the much smaller cosmological horizon 
scale at the time of the phase transition.

The two point correlator of a stochastic, homogeneous, isotropic magnetic field 
contains two independent functions. In physical space, these
are the normal and helical correlation functions ($M_N(r)$ and $M_H(r)$); in momentum space, these are the symmetric and antisymmetric power spectra ($S(k)$ and $A(k)$). 
There are several tools to measure the normal correlator (or the symmetric
power spectrum) of cosmological magnetic fields
but measuring helicity directly is more challenging.  The detection
of helicity of cosmological magnetic fields has been a primary motivation for this
work, as few other direct schemes to detect helicity have been proposed~\cite{Kahniashvili:2005yp}.

A measurement of the normal correlator itself, together with some theoretical
input, provides indirect access to the helicity of the magnetic field. 
This is because the presence of helicity influences the evolution of the normal 
correlator, and the exponents
characterizing the normal correlator can give us some information about the 
magnetic helicity~(for the evolution in the cosmological context,
see~Ref.\cite{Christensson:2002xu,Campanelli:2004wm,Banerjee:2004df,Campanelli:2007tc,Boyarsky:2011uy,Kahniashvili:2012uj}).
The shape of the normal correlator, in particular the existence 
of a peak in the distribution can also inform us about the epoch of magneto-genesis. 
If magnetic fields are produced during inflation, the symmetric power spectrum can 
be expected to be scale invariant with significant power on the present day horizon 
scale. 

Direct measures of the helical power spectrum are more difficult.
For {\it astrophysical} magnetic helicity, it has been suggested that the correlation 
between Faraday rotation measurement and the polarization degree of radio synchrotron
emission~\cite{2010JETPL..90..637V,Junklewitz:2010ux,Oppermann:2010uy} can be
used.  For {\it cosmological} magnetic helicity, Ref.~\cite{Kahniashvili:2005yp} has shown 
that correlations in the arrival momenta of very high energy cosmic rays are sensitive to
the intervening magnetic helicity provided the cosmic ray source locations are
known.  Cosmic Microwave Background (CMB) anisotropy observations may also 
permit a measure of the helicity
through  non-vanishing cross-correlation between the temperature and B-mode 
polarization anisotropies and between the E-mode and B-mode polarization 
anisotropies~\cite{Caprini:2003vc,Kahniashvili:2005xe,Kunze:2011bp}. 

In the present paper, we propose a scheme to measure both the normal and the
helical correlation functions of cosmological magnetic fields by using
TeV blazar observations.  Gamma rays with energy greater than $\sim 1~{\rm TeV}$
can scatter with ambient ``extragalactic background light'' (EBL) 
photons to pair produce electrons and positrons. 
The generated electrons and positrons create a secondary cascade of GeV 
gamma rays through the Inverse Compton (IC) scattering of CMB photons.  As the
electron and positron trajectories are bent due to the Lorentz force by a 
magnetic field, the GeV photon cascade carries information about the structure 
of the cosmological magnetic field.
We show that cross-correlations between the arrival directions of the
secondary cascade gamma rays at different energies are related to the
correlation function of the extragalactic magnetic
fields.  If we imagine the cascade
photon arrival direction to be a vector in the 
plane of observation, the inner product of vectors at different energies gives 
the normal part of
the magnetic field correlation, and the outer product gives the helical correlator.

In Sec.~\ref{geometry} we describe the basic geometry of the process and
evaluate the arrival direction of cascade
photons as a function of the intervening
magnetic field. In Sec.~\ref{correlators} we evaluate correlators of the arrival
directions of cascade 
photons and relate them to the magnetic field correlation
functions. Our results are placed in the context of observations in Sec.~\ref{observations}
where we discuss estimators for the theoretical correlation functions we have
found in Sec.~\ref{correlators}. Our analysis uses many simplifying assumptions 
that we discuss together with conclusions in Sec.~\ref{conclusions}.
Throughout this paper, we use natural units: $\hbar=1=c$.

\section{Geometrical setup and deflection angle}
\label{geometry}

TeV gamma rays from distant sources at redshift $z_s$ cannot propagate 
freely over cosmological distances because, at such energies, interaction with 
the EBL can produce electrons and positrons. The mean free path of a
gamma ray with energy $E_{\rm TeV}$  is given by \cite{Neronov:2009gh}
\begin{equation}
 D_{\rm TeV} (E_{\rm TeV}) \sim 80 { \kappa \over (1+z_s)^2}  ~{\rm Mpc}~\left( {E_{\rm TeV} \over 10~{\rm TeV}}\right)^{-1},
 \label{eq:tev-meanfree}
\end{equation}
where $\kappa$ is a numerical factor which accounts
for the model uncertainties of EBL. Here we take $\kappa \sim 1$~\cite{Neronov:2009gh}.
Electrons and positrons generated by the TeV gamma ray lose
energy by the production of a secondary gamma ray cascade through the IC
scattering of CMB photons, and have a mean free path,
\begin{equation}
 D_e \sim 
 30 ~{\rm kpc} ~ (1+z_e)^{-4} \left( {E_e \over 10~{\rm TeV} }\right)^{-1},
\end{equation}
where $z_e$ is the typical redshift at which TeV gamma rays create pairs, and 
$E_e  \sim E_{\rm TeV}/2$ is the electron energy. The up-scattered CMB photon has energy, 
\begin{equation}
 E_\gamma = {4 \over 3} (1+z_e)^{-1} \epsilon_{CMB} \left( {E_e \over m_e }\right)^2
  \sim 88 ~{\rm GeV} ~\left( {E_{\rm TeV} \over 10~{\rm TeV} }\right)^{2},
  \label{eq:gamma-ray}
\end{equation}
where $\epsilon_{\rm CMB}= 6 \times 10^{-4} (1+z_e)~{\rm eV}$ is the typical energy 
of CMB photons.
Therefore we observe cascade
gamma rays with energy $\sim E_\gamma $,
in addition to TeV gamma rays with energy $\sim E_{\rm TeV}$.

\begin{figure}
  \begin{center}
   \includegraphics[width=100mm]{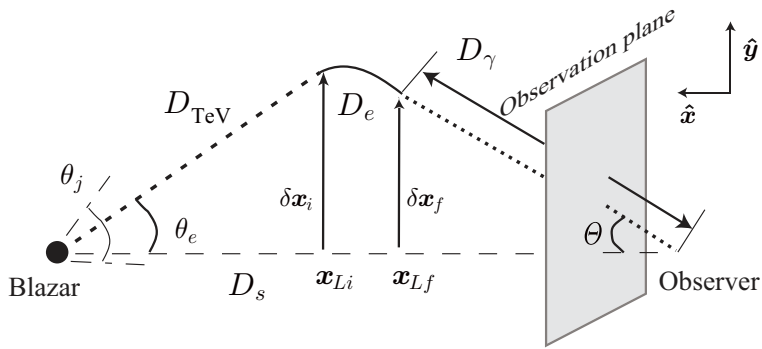}
  \end{center}
\caption{The blazar on the left beams TeV photons within a jet of opening angle $\theta_j$.
The observer is most likely located at the edge of the jet, not on the axis. TeV photons 
pair produce after propagating a distance $D_{\rm TeV}$. The pairs are bent by
ambient magnetic fields and up-scatter CMB photons that propagate a distance 
$D_\gamma$ to the observer. The emission angle $\theta_e$, the
observation direction $\Theta$, the distance to the source $D_s$, and the
pair creation and IC scattering event positions, $(x_{Li},\delta x_i)$, 
$(x_{Lf},\delta x_f)$ are also shown.}
\label{fig:tri}
\end{figure}

It is believed that TeV gamma rays are beamed from distant blazars in a narrow jet
of opening angle $\theta_j \sim 5 ^\circ ~[\Gamma /10]^{-1}$ where $\Gamma$
is the Lorentz factor of the gamma ray emitting plasma. The highest energy photons
are observed if we lie within the opening angle of the jet. Denoting the angle between
the source direction and the jet axis by $\theta_o$, we require $\theta _o < \theta_j$.
As emphasized in Ref.~\cite{1991ApJ...382..501U}, the most likely situation is that we are located
on the edge of the cone, as depicted in Fig.~\ref{fig:tri}.

Suppose that the blazar is located at ${\bm x}_s = D_s \hat {\bm n}_s $ 
in the observer 
frame. We will assume for simplicity that the redshift of
the source is less than one and take $1+z_s \sim 1$.
If there are no magnetic fields in the IGM, we will observe cascade 
gamma rays
due to TeV gamma rays also in the direction $\hat {\bm n}_s$.
However if cosmological magnetic fields are present, the cascade 
gamma rays
arrive from different directions. We now evaluate the arrival direction
assuming a stochastic, homogeneous, and isotropic magnetic field.

Consider an observed cascade 
gamma ray that resulted from a TeV gamma ray
with energy $E_{\rm TeV}$ that was emitted at $t=0$ at an angle $\theta_e$ from the
line of sight as depicted in Fig.~\ref{fig:tri}.
The TeV gamma ray produces an electron at the position ${\bm x}_i$ at
time $t=t_i$, where $t_i  = D_{\rm TeV}$.
The produced electron has momentum ${\bm P}_i $
whose amplitude $P_i$ corresponds to the energy $E_e \approx E_{\rm TeV}/2$.
Since the opening angle of the electron-positron pair is very small, of order
$m_e /E_e \sim 10^{-7}$,
the direction ${\bm P}_i$ is the same as the direction of the initial TeV 
gamma ray.

The momentum of the electron changes on propagation due to the Lorentz 
force,
\begin{equation}
 {\bm P}(t) = {\bm P}_i + q \int_{t_i}^t dt' ~ {\bm v }(t') \times {\bm B}({\bm x}(t')) 
\label{eq:momentum_evo}
\end{equation}
where $q=\pm e$ is the electron/positron charge, ${\bm x}(t) $ and ${\bm v }(t)$ 
are the position and velocity of the electron (or positron), namely 
${\bm v}(t) = \dot {\bm x}(t)$ where the overdot denotes differentiation with respect 
to time. (For convenience, from now on we shall refer to the charged particle as being 
the electron.)

We will now decompose all vectors in components parallel
and perpendicular to the source direction.
For example, the momentum and the position of the electron at time $t$ is decomposed as
\begin{equation} 
{\bm P}(t) = {\bm P}_{L} (t ) +\delta {\bm p} (t),
 \qquad
 {\bm x}(t) = {\bm x}_{L} (t ) +\delta {\bm x} (t),\label{eq:decompose}
\end{equation}
where the subscript ${L}$ means the component parallel to the source direction
(line-of-sight for TeV source).
Therefore, the vector $\delta {\bm p} (t)$ and $\delta {\bm
x} (t)$ are the deviations induced by the magnetic field.

In terms of the decomposed components,
Eq.~(\ref{eq:momentum_evo}) can be written as
\begin{equation}
 {\bm P}_{L}(t) +\delta {\bm p}(t) = {\bm P}_{Li} +\delta {\bm
  p}_{i} +
 q \int_{t_i}^t dt' ~ [{\bm v}_{L} (t' ) +\delta {\bm v} (t')] \times
 {\bm B}({\bm x}(t ')) ,
\label{eq:momentum_evo_m}
\end{equation}
where ${\bm P}_{Li} $ and $\delta {\bm  p}_{i}$ are the momentum
components at time $t=t_i$. Note that at this stage, instead of replacing 
${\bm x}$ by ${\bm x}_L$ in the argument of ${\bm B}$, we perform the integration 
along the actual path ${\bm x}(t)$. This is important if the magnetic fields have
significant power on small scales, {\it i.e.}, a blue spectrum.

The bending angle of the electron is estimated as 
$\delta = D_e /R_L \sim 1.2 \times 10^{-3} [B/10^{-16} ~{\rm G}]
[E_{\rm TeV}/10~{\rm TeV}]^{-2}$ where $R_L = E_e/qB$ is the Larmor radius. 
Here we have assumed a magnetic field coherence scale larger than $D_e$;
otherwise the electron trajectory would be diffusive yielding a smaller estimate
for $\delta$.
Then the maximum deviation from the source direction is 
$\delta {x}_i \sim 90 ~{\rm kpc} (1-D_{\rm TeV}/D_s) [B/10^{-16} ~{\rm G}]
[E_{\rm TeV}/10~{\rm TeV}]^{-3}$. Since the bending angle is small, we can
treat  $\delta {\bm p}$, $\delta {\bm x}$ and $\delta \bm v$ as perturbations.
To linear order in the magnetic field strength, Eq.~(\ref{eq:momentum_evo_m}) becomes
\begin{equation}
 \delta {\bm p}(t) = \delta {\bm p}_{i} +
 q \int_{t_i}^t dt' ~ {\bm v}_{L} (t ') \times 
{\bm B}({\bm x} (t' )) .
\label{eq:d_momentum_evo}
\end{equation}

The electron energy $E_e$ is constant during this process, since a
magnetic field does no work. Dividing Eq.~(\ref{eq:d_momentum_evo}) by
$E_e$, we obtain the velocity,
\begin{equation}
 \delta {\bm v}(t) = \delta {\bm  v}_{i} +
 {q \over E_e} \int_{t_i}^t  dt' ~ {\bm v}_{L} (t' ) \times {\bm B}({\bm x} (t' )) ,
\label{eq:velocity_evo_m}
\end{equation}
and another integration gives the trajectory,
\begin{equation}
 \delta {\bm x}(t) -\delta {\bm x}_i = \delta {\bm  v}_{i} (t-t_i) +
 {q \over E_e}  \int^{t}_{t_i} dt'' \int ^{t''} _{t_i}dt' ~ {\bm v}_{L} (t' ) \times {\bm B}({\bm x} (t' )) .
\label{eq:position_evo_m}
\end{equation}

At any point in the electron's trajectory there is a probability of an IC scattering event 
with a CMB photon. We simplify the present analysis by assuming
that the electron travels a fixed distance $D_e (E_e)$, and at final time $t_f$ 
up-scatters a CMB photon. Since the electron is ultra-relativistic with Lorentz boost
factor $E_e/m_e \sim 10^7$,
the cascade 
gamma ray propagates along the momentum of the electron.
Therefore the arrival direction of the cascade 
gamma ray
is the same as the direction of the position ${\bm x}_f ={\bm x}(t_f)$.

We now define the vector ${\bm \Theta}$ in the observation plane by
\begin{equation}
{\bm  \Theta} \equiv \frac{\delta {\bm x}_i - \delta {\bm x}_f}{D_e}
\label{eq:theta_obs}
\end{equation} 
where $\delta {\bm x}_f $ is $\delta {\bm x} (t)$ at $t=t_f$.
The magnitude, $|{\bm \Theta}|$, is the observed angle $\Theta$ shown
in Fig.~\ref{fig:tri}, and
the direction in the observation plane corresponds to the azimuthal direction
of the original TeV gamma ray. 

We will now relate ${\bm \Theta}$ to the magnetic field using Eq.~(\ref{eq:position_evo_m}).
However $\delta {\bm  v}_{i}$ is still unknown in Eq.~(\ref{eq:position_evo_m}). 
To evaluate it, we first need to express the emission angle $\theta_e$ in
terms of the observed angle $\Theta$. 
Applying the trigonometric law of sines to the triangle formed by the source,
observer and electron position -- recall that $D_e \ll D_s$ -- we obtain
\begin{equation}
\theta_e \approx  {D_\gamma \over D_{\rm TeV}} \Theta 
\approx  \frac{D_s -D_{\rm TeV}}{D_{\rm TeV}} \Theta .
\end{equation}
In terms of ${\bm \Theta}$,
the vector $\delta {\bm v}_i$ can now be written as
\begin{equation}
\delta {\bm v}_i = v_e  \theta_e {\hat {\bm y}} = v_e {D_s-D_{\rm TeV} \over D_{\rm TeV}} {\bm \Theta},
\label{eq:vi}
\end{equation}
where $v_e$ is the magnitude of the electron velocity, and ${\hat {\bm y}}$ is 
the unit vector perpendicular to the line-of-sight (see Fig.~\ref{fig:tri}).

Now Eqs.~(\ref{eq:position_evo_m}), (\ref{eq:theta_obs}) and (\ref{eq:vi}) give,
\begin{equation}
{\bm \Theta} (E_\gamma) = - {q D_{\rm TeV} \over E_e D_e D_s}  
                  \int^{t_f}_{t_i} dt'' \int ^{t''} _{t_i}dt' ~ {\bm v}_{L} (t' ) \times {\bm B}({\bm x} (t' )) ,
\label{eq:final_ang}
\end{equation}
where we can rewrite $D_{\rm TeV}$, $D_e$, and $E_e$ as functions of
$E_\gamma$,
\begin{equation}
 D_{\rm TeV} (E_{\rm TeV}) \sim 80 ~{\rm Mpc}~\left( {E_\gamma \over 88~{\rm GeV}}\right)^{-1/2},
 \label{eq:tev-meanfree-gev}
\end{equation}
\begin{equation}
 D_e \sim 
 30 ~{\rm kpc} ~ \left( {E_\gamma \over 88~{\rm GeV} }\right)^{-1/2},
\end{equation}
\begin{equation}
 E_e 
  \sim 10 ~{\rm TeV} ~\left( {E_{\gamma} \over 88~{\rm GeV} }\right)^{1/2}.
  \label{eq:gamma-ray-gev}
\end{equation}

With these relations, the magnitude of ${\bm \Theta} (E_\gamma)$ is roughly estimated as
\begin{equation}
 \Theta (E_\gamma) \approx
  {q D_{\rm TeV}  D_e \over E_e D_s}  ~{v}_{L} {B}
 \approx
 7.3 \times 10^{-5} ~ \left( { B  \over 10^{-16} ~{\rm Gauss}}\right) \left( {E_\gamma
  \over 100 ~{\rm GeV}}\right)^{-3/2} \left( {D_s
  \over 1000 ~{\rm Mpc}}\right)^{-1} .
\label{eq:final_ang_est}
\end{equation}

\section{Correlators and predictions}
\label{correlators}

The vector ${\bm \Theta}(E_\gamma)$ describes the position of the
observed cascade 
gamma ray in the
observation plane but its magnitude depends on geometrical factors such as the distance to
the source (Eq.~(\ref{eq:final_ang})). We define a rescaled vector to remove such dependence,
\begin{equation}
 {\bm Q}(E_\gamma) \equiv  {E_e D_s \over q D_{\rm TeV} D_e} {\bm \Theta} (E_\gamma) .
\end{equation}

Now we are interested in two types of correlators of the ${\bm Q}(E_\gamma)$ vectors
\begin{equation}
 F(E_1, E_2) = \langle {\bm Q} (E_1) \cdot {\bm Q} (E_2) \rangle,
 \label{Fdef}
\end{equation}
\begin{equation}
 G(E_1, E_2) = \langle {\bm Q} (E_1)  \times {\bm Q} (E_2) \cdot \hat {\bm x} \rangle,
\label{Gdef}
\end{equation}
where $E_1$ and $E_2$ are two energies of the observed cascade 
gamma rays, and we
recall that we have set up our coordinate system so that the $x-$axis is along the line
of sight, {\it i.e.}, $\hat {\bm x}$ is normal to the observation plane.

The ensemble value $\langle Q_i (E_1) Q_{i'} (E_2) \rangle$, where $i$
and $i'$ denote components in the observation plane, can be written by using the magnetic field correlation function, Eq.~(\ref{eq:final_ang}),
\begin{equation}
 \langle Q_i (E_1) Q_{i'} (E_2) \rangle
=
\epsilon_{ijl} \epsilon_{i'j'l'} v_L^j (E_1 ) v_L^{j'} (E_2)
\int_{t_{1 \rm i}} ^{t_{1\rm f}} {dt_1' \over D_{e1}}
\int_{t_{1 \rm i}} ^{t'_1} {dt_1 \over D_{e1}}~
\int_{t_{2 \rm i}} ^{t_{2 \rm f}} {dt_2' \over D_{e2}}
\int_{t_{\rm i}} ^{t'_2} {dt_2 \over D_{e2}}~
\langle
B^l ({\bm x}(t_1,E_1))
B^{l'} ({\bm x}(t_2,E_2))
\rangle
,
\label{eq:ensemble}
\end{equation}
where $D_{e1}$ and $D_{e2}$ are $D_e$ 
for the cascade 
gamma ray with energy $E_1$ and $E_2$ respectively.
Note that the velocity ${\bm v}_L = (1,0,0)$ 
and $t_{\rm f}- t_{ \rm i} = D_{e}$. 

The correlation function of a stochastic, homogeneous, and isotropic magnetic field is
given by~\cite{1975mit..bookR....M}
\begin{equation}
 \langle B_i ({\bm x} +{\bm r}) B_j ({\bm x})  \rangle =
  M_N(r)  \left[ \delta_{ij} -{r_i r_j \over r^2} \right]+
M_L(r)  {r_i r_j \over r^2} + M_H (r) \epsilon_{ijl} r^l,
\label{BBcorr}
\end{equation}
where $M_N(r)$, $M_L(r)$, and $M_H(r)$ are the correlation
functions for the normal, longitudinal, and helical parts of the
magnetic fields. Due to the homogeneity and isotropy of the magnetic fields, these
correlations depend only on the separation distance $r = |\bm r|$.
The divergence-less condition gives
\begin{equation}
 M_N(r) ={1 \over 2 r} {d \over dr} (r^2 M_L(r)).
\end{equation}

The stochastic fields are often described by power spectra in
Fourier space. The magnetic field correlation function in Fourier space is
\begin{equation}
 \langle \tilde{B}^* _i ({\bm k}) \tilde{B} _j ({\bm k}') \rangle
= (2 \pi)^3 \delta^3 ({\bm k} -{\bm k}') 
\left[ \left(\delta _{ij} - {k_i k_j \over k^2} \right) 
 S (k)  + i \epsilon_{ijl} {k_l \over k}
 A(k) 
\right],
\end{equation}
where $S(k)$ and $A(k)$ are the symmetric and antisymmetric (helical) parts of the
magnetic field power spectrum. The functions $S(k)$ and $A(k)$ are related to the 
correlation functions $M_N(r)$, $M_L(r)$, and $M_H(r)$ as in~\cite{1975mit..bookR....M}.

Therefore the ensemble average, Eq.~(\ref{eq:ensemble}), can also be
decomposed into three parts: the normal, longitudinal, and helical parts,
\begin{equation}
 \langle Q_{i} (E_1) Q_{i'} (E_2) \rangle
= C_{N i i'}  + C_{L i i'}  +C_{H i i'} .
\end{equation}
with each of the terms given by four integrations as in Eq.~(\ref{eq:ensemble}).
The separation scale, $r$, appearing in the correlators $M_N(r)$, $M_L(r)$ and
$M_H(r)$ is the magnitude of the separation vector
\begin{equation}
 {\bm r} (t_1,t_2,E_1,E_2) \equiv  {\bm x} (t_1,E_1) -{\bm x} (t_2,E_2) .
 \end{equation}
Since cosmological magnetic fields are weak, this can be approximated as
\begin{equation}
 {\bm r} (t_1,t_2,E_1,E_2) \approx {\bm x} _L (t_1,E_1) -{\bm x} _L (t_2,E_2).
\end{equation}
Therefore,
\begin{equation}
 {\bm r} (t_{1 },t_{2 },E_1,E_2) = ( D_{\rm TeV} (E_1) -D_{\rm TeV}(E_2) +
 (t_1-t_{1 \rm i}) -(t_2-t_{2 \rm i})) {\hat {\bm x}} .
\end{equation}
For $t_1=t_{1 \rm  f}$ and $t_2=t_{2 \rm  f}$, the separation scale becomes
\begin{equation}
 r (t_{1 \rm f},t_{2 \rm f},E_1,E_2) = D_{\rm TeV} (E_1) -D_{\rm TeV}(E_2) +D_e (E_1) -D_e(E_2).
\end{equation}

In the case with $i = i' \neq x$, $ C_{L i i}$ and $C_{H i i} $
vanish, because only the $\hat {\bm x}$ components of ${\bm v}_L$ and ${\bm r}$ 
are non-vanishing. However the normal correlator does not vanish and
$ C_{N i i}$ is given by
\begin{equation}
 C_{N i i} (E_1, E_2) = 
\int_{t_{1 \rm i}} ^{t_{1\rm f}} dt_1' \int_{t_{1 \rm i}} ^{t'_1} dt_1~
\int_{t_{2 \rm i}} ^{t_{2 \rm f}} dt'_2 \int_{t_{\rm i}} ^{t'_2} dt_2~
M_N( r (t_{1 },t_{2 },E_1,E_2)).
\label{eq:CN_ij}
\end{equation}

In the case with $i \neq i'$, $i \neq x$ and $i' \neq x$,
while $ C_{N i i'}$ and $C_{L i i'} $
vanish, $ C_{H i i'}$ has non-zero value,
\begin{equation}
C_{H i i'} =
\epsilon_{ii'z}
\int_{t_{1 \rm i}} ^{t_{1\rm f}} dt_1' \int_{t_{1 \rm i}} ^{t'_1} dt_1~
\int_{t_{2 \rm i}} ^{t_{2 \rm f}} dt'_2 \int_{t_{2 \rm i}} ^{t'_2} dt_2~
M_H(r(t_1,t_2,E_1,E_2)) ( {\bm r} (t_1,t_2,E_1,E_2) \cdot \hat {\bm x}),
\label{eq:CH_ij}
\end{equation}

Therefore $F$ and $G$ can be written directly in terms of the
magnetic field correlation functions,
\begin{equation}
 F(E_1, E_2) =  2 
\int_{t_{1 \rm i}} ^{t_{1\rm f}} {dt_1' \over D_{e1}} \int_{t_{1 \rm i}}
^{t'_1} {dt_1 \over D_{e1}}~
\int_{t_{2 \rm i}} ^{t_{2 \rm f}} {dt'_2 \over D_{e2}} \int_{t_{\rm i}}
^{t'_2} {dt_2 \over D_{e2}}~
M_N( r (t_{1 },t_{2 },E_1,E_2)),\label{eq:f}
\end{equation}
\begin{equation}
 G(E_1, E_2) =  2
  \int_{t_{1 \rm i}} ^{t_{1\rm f}} {dt_1' \over D_{e1}} \int_{t_{1 \rm i}}
^{t'_1} {dt_1 \over D_{e1}}~
\int_{t_{2 \rm i}} ^{t_{2 \rm f}} {dt'_2 \over D_{e2}} \int_{t_{\rm i}}
^{t'_2} {dt_2 \over D_{e2}}~
M_H(r(t_1,t_2,E_1,E_2)) ( {\bm r} (t_1,t_2,E_1,E_2) \cdot \hat {\bm x}).
\label{eq:g}
\end{equation}
It is worth pointing out that $F(E_\gamma ,E_\gamma) \ne 0$ but
$G(E_\gamma,E_\gamma) =0$.

We have already seen that $D_{\rm TeV} \gg D_e $ independently of $E_\gamma$, 
and so the separation scale in the magnetic field correlation function can be approximated as,
\begin{equation}
  r (t_{1 },t_{2 },E_1,E_2) \approx D_{\rm TeV} (E_1) -D_{\rm TeV}(E_2).
  \label{rE1E2}
\end{equation}
Then recalling $t_{\rm f} -t_{\rm i} = D_e$, we can do the integrations in 
Eqs.~(\ref{eq:f}) and (\ref{eq:g}) to get
\begin{equation}
 F(E_1, E_2) \approx  {1 \over 2} 
M_N( |r_{12}| )
,\label{eq:f-2}
\end{equation}
\begin{equation}
 G(E_1, E_2) \approx  {1 \over 2} 
M_H( |r_{12}| ) r_{12},
\label{eq:g-2}
\end{equation}
where $r_{12}$ is $ r_{12} = D_{\rm TeV} (E_1) -D_{\rm TeV}(E_2)$.
This shows that we can obtain both the non-helical and the helical magnetic field correlation 
functions through the correlators $F(E_1, E_2)$ and $ G(E_1, E_2)$ 
constructed from the arrival information
of cascade photons. This is the main result of this paper.

\begin{figure}
  \begin{center}
   \includegraphics[width=100mm]{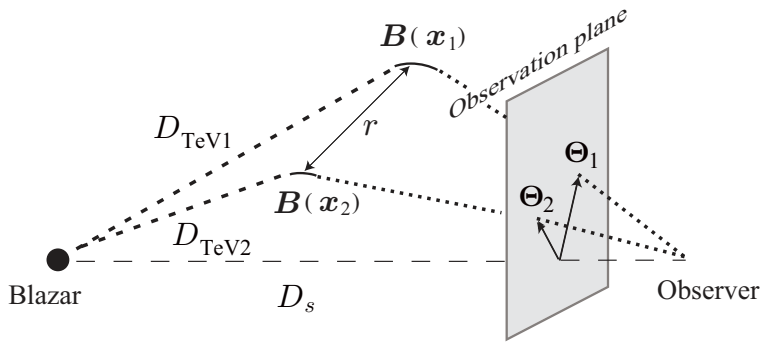}
  \end{center}
\caption{Events at two different energies sample the magnetic field
in regions of size $D_e \sim 30~{\rm kpc}$ (solid lines at the vertices
of the triangles). The regions themselves are separated by
distance $r$ which can be $\sim 100~{\rm Mpc}$ depending on the
energy difference of the two events. Energy resolution of the detector
translates into a lower limit on the separation at which the magnetic
field correlations can be probed.
 }
\label{fig:cor}
\end{figure}

Note, as shown in Fig.~\ref{fig:cor}, that the obtained correlator is on 
the scale $r_{12}$ which can be much
larger than the scale $D_e$. This is because the magnetic field gets 
correlated at the spatial points where pair production occurs and these can
be separated by hundreds of Mpc depending on the values chosen for
$E_1$ and $E_2$. It remains an interesting open question if further
refinements of the above method can lead to correlators on scales
smaller than $D_e \sim 30~{\rm kpc}$. 

 \section{Observations and estimators}
 \label{observations}
 
In the previous section, we have considered correlators in the arrival directions
of cascade 
photons and expressed them in terms of correlators of the intervening
magnetic field. In this section we consider the problem of determining magnetic
field correlators from the observers point of view: given cascade 
arrival direction
data, what quantity should be calculated that corresponds to the magnetic
field correlator? The calculated quantity will at best be an estimator for the
magnetic field correlator because observations are limited in number, while
the magnetic field correlator in Eq.~(\ref{BBcorr}) is over an infinite ensemble
of realizations.

In a more realistic setting -- see simulations in Ref.~\cite{Elyiv:2009bx} --
observed gamma rays with some energy $E$
are expected to be scattered around a typical observation direction
on the observational plane. 
The scattering is due to differences in 
the magnetic field along the trajectories of different electrons and also
the stochasticity of the EBL and CMB photons.
We have not considered the latter fluctuations in this paper but the variation 
in the magnetic field can be smoothed out by taking the average
$\langle {\bm \Theta} (E) \rangle$ over all observed gamma rays with 
the same energy $E$. 
Even if the magnetic fields have small-scale structure, {\it i.e.},
a blue spectrum, the averaging procedure should yield the correct magnetic 
field correlator on scales larger than the smoothing scale, 
$|\delta {\bm x}| \sim 90~{\rm kpc}$.

There is a second way to smooth out fluctuations in the magnetic
field. This is by realizing that the correlation functions $F(E_1,E_2)$ and
$G(E_1,E_2)$ evaluated in the previous section depend on only one
function, $r(t_1,t_2,E_1,E_2)$. Hence there are many choices
of $E_1$ and $E_2$ yielding the same $r$, and one of the energy
variables can be integrated out. This corresponds to averaging over all 
pairs of energies such that the distance $r$ remains fixed (see Fig.~\ref{fig:cor}). Using
Eq.~(\ref{rE1E2}) we see that $r(t_1,t_2,E_\gamma,E_\gamma+\delta E)$ 
is independent of $E_\gamma$ provided
\begin{equation}
\delta E (E_\gamma , r) = {27 ~{\rm GeV}} \left[
\left( {r \over 10 ~{\rm Mpc}} - \sqrt{88 ~{\rm GeV} \over E_\gamma}
\right)^{-2} -{88 ~{\rm GeV} \over E_\gamma} \right].
\end{equation}
We can now average Eqs.~(\ref{eq:f-2}) and (\ref{eq:g-2}) over $E_1$
while taking $E_2=E_1+\delta E$,
\begin{equation}
M_N( r ) \approx 2 \int  {d E_\gamma \over \Delta E}
 {\bm Q}(E_\gamma ) \cdot {\bm   Q}(E_\gamma + \delta E (E_\gamma ,r)),
\end{equation}
\begin{equation}
r  M_H( r )  \approx 2  \int  {d E_\gamma \over \Delta E}
 {\bm Q}(E_\gamma ) \times {\bm   Q}(E_\gamma + \delta E (E_\gamma ,r))
  \cdot \hat {\bm x},
\end{equation}
where $\Delta E$ is the integration range of the observation energy. The
vector ${\bm Q}(E)$ now denotes the {\it average} (rescaled) direction 
vector for cascade photons with energy $E$.

A third way to perform the ensemble average is to use observations
of many TeV blazars because cascade 
gamma rays from different blazars 
sample magnetic fields along different path. Hence we can obtain the 
magnetic field correlation function using
\begin{equation}
M_N (r) \approx
{2 \over  N} \sum_{\alpha } ^N  \left[
2 \int  {d E_\gamma \over \Delta E}
 {\bm Q}(E_\gamma ) \cdot {\bm   Q}(E_\gamma + \delta E (E_\gamma ,r))
\right]_\alpha,
\label{eq:observe_ave1}
 \end{equation}
 \begin{equation}
r  M_H (r) \approx
 {2 \over  N}\sum_{\alpha } ^N \left[
2  \int  {d E_\gamma \over \Delta E} 
 {\bm Q}(E_\gamma ) \times {\bm   Q}(E_\gamma + \delta E (E_\gamma ,r))
  \cdot \hat {\bm x}
 \right]_\alpha,
\label{eq:observe_ave}
\end{equation}
where $\alpha$ labels the blazar.

Finally we would like to remark that the most likely observation of a TeV
blazar is when we are positioned at the edge of the 
jet~\cite{1991ApJ...382..501U}, but
we are even more likely to be located outside the jet opening angle. Then 
we will not observe the TeV source but will still receive some cascade GeV 
photons. In this case, if there is reason to suppose that the observed 
photons are indeed from a cascade, we can extend our correlator by 
replacing ${\bm Q}(E)$ in Eqs.~(\ref{eq:observe_ave1}), (\ref{eq:observe_ave}) 
by ${\bm Q}(E)-{\bm Q}(E_*)$ where $E_*$ is the highest energy (and least
deviated) cascade photon that is observed. We can still draw some conclusions 
about the magnetic field correlators though the analysis is more involved.

\section{Discussion and conclusions}
\label{conclusions}

Our main result is the connection between cosmological magnetic field
correlators and the correlators of cascade photons as given in 
Eqs.~(\ref{eq:f-2}) and (\ref{eq:g-2}) together with the definitions
in Eqs.~(\ref{Fdef}) and (\ref{Gdef}). 
In the observational context, this
connection can be written as in Eqs.~(\ref{eq:observe_ave1}) and
(\ref{eq:observe_ave}). These relations suggest that observations of
cascade photons may be used to directly study the non-helical and
helical spectra of cosmological magnetic fields. There are few other
direct ways to probe the helicity of cosmological magnetic fields and
this is an important feature of the technique we have described.

Our analysis will need further development as it has used many
simplifying assumptions. For example, the stochastic interactions of the
TeV photons with the EBL photons will lead to a probability distribution for
$D_{\rm TeV}$ while we have used a fixed value. Similarly there will be
probability distributions for the other interactions, and the magnetic
field will have some structure on small scales. In addition, if the
blazar is at high redshift, cosmological expansion will become
important. These effects can best be studied using Monte Carlo
simulations similar to the ones described in
Refs.~\cite{Dolag:2009iv,Elyiv:2009bx}.  The simulations should be able
to confirm if the magnetic field correlators can be recovered from
realistic data, which hopefully will become available in the near future.

\acknowledgments
We thank Nat Butler and Francesc Ferrer for discussion. This work was 
supported by the DOE at ASU. 


\end{document}